\documentclass[aps,twocolumn,showpacs]{revtex4}
\usepackage{graphicx}
\usepackage{epsfig}
\usepackage{times}
\newcommand{\be}{\begin{equation}}
\newcommand{\ee}{\end{equation}}
\newcommand{\bea}{\begin{eqnarray}}
\newcommand{\eea}{\end{eqnarray}}

\begin{document}
\title{Role of loop entropy in the force induced melting of DNA hairpin}
\author{Garima Mishra$^{1}$, Debaprasad Giri$^{2}$, M. S. Li$^{3}$ 
and Sanjay Kumar$^{1}$}
\affiliation{$^1$Department of Physics, Banaras Hindu University,
Varanasi 221 005, India \\
$^2$Department of Applied Physics, IT, Banaras Hindu University,
Varanasi 221 005, India \\
$^3$ Institute of Physics, Polish Academy of Sciences, Al. Lotnikow
32/46, 02-668 Warsaw, Poland 
}
\date{\today}

\begin{abstract}
Dynamics of a single stranded DNA, which can form a hairpin have 
been studied in the constant force ensemble. Using Langevin dynamics 
simulations, we obtained the force-temperature diagram, which differs 
from the theoretical prediction based on the lattice model. Probability 
analysis of the extreme bases of the stem revealed that at high temperature, 
the hairpin to coil transition is entropy dominated and the loop contributes
significantly in its opening. However, at low temperature, the transition is 
force driven and the hairpin opens from the stem side. It is shown that the 
elastic energy plays a crucial role at high force. As a result, the phase 
diagram differs significantly with the theoretical prediction.

\end{abstract}
\maketitle

\section{Introduction}
In recent years, considerable experimental, theoretical and numerical efforts
have been made to understand the dynamics and kinetics of DNA/RNA hairpin 
\cite{wallace,bonnet,Goddard,zhang,kim,nivon,lin,kenward,errami}. 
This is because 
the hairpin participates in many biological functions {\it e.g.} replication, 
transcription, recombination, protein recognition, gene regulation 
and in understanding the secondary structure of  RNA molecules \cite{Zazopoulos,
Froelich,Trinh,Tang}. Moreover it has been used as a tool in the form of 
Molecular Beacon, which provides increased specificity of target recognition in 
DNA and RNA \cite{bonnet,Goddard,tan,broude,ysli,drake}.  

The hairpin is made up of a single stranded DNA (ssDNA)/RNA,  which carries 
sequence of  bases that are complementary to each other in each of its two 
terminal regions.  When the base pairing of these two remote 
sequences are formed,  it gives rise to the structure of hairpin, consisting 
of two segments: a stem, which comprises of a short segment of the DNA helix and, 
a loop of single strand carrying the bases that are not paired (Fig. 1).
Structures of the hairpin  are not static, as they fluctuate among many
different conformations. Broadly speaking, all of these conformations may 
be classified into the two states: (i) the open state and (ii) the closed state 
(Fig. 1). The open state has high entropy because of the large number of 
conformations accessible to the ssDNA, whereas  the closed state is 
a low-entropy state, where enthalpy is involved in the
base pairing of the stem. It was shown that the closed-to-open transition 
requires a sufficiently large amount of energy to unzip all of the base pairs of the 
stem, whereas the open-to-close transition involves a lower energy barrier 
\cite{bonnet,Goddard,Chen,asain}.
The thermodynamics and kinetics of the hairpin are well studied \cite{bonnet,Goddard}.
The melting temperature ($T_m$) of the hairpin, at which the stem is denaturated and the 
hairpin behaves like a single polymer chain,  is measured by Fluorescence 
Correlation Spectroscopy (FCS) and in good agreement with theory \cite{bonnet,Goddard,Chen}.  
It was also found that the rate of closing strongly  depends on the 
properties of the hairpin loop, such as length and rigidity, whereas  the rate of 
opening remains relatively insensitive to these properties 
\cite{bonnet,Goddard,Chen}.

Single molecule force measuring techniques can manipulate biomolecules 
(DNA, protein etc) by applying a force on the pico-newton (pN) 
scale \cite{Bockelmann,smith,ubm,kumarphys,woodside}. These 
techniques such as, atomic force microscope, optical tweezers, magnetic tweezers etc.  
have enhanced our understanding about the structural and functional properties of 
biomolecules and shed important information about the molecular forces involved in 
the stability of biomolecules \cite{Bockelmann,smith,ubm,kumarphys}. 
In the case of the DNA hairpin, if the applied force is close to a critical value, then 
the hairpin fluctuates between the closed and open state. Therefore,  efforts 
were  mainly made to understand the kinetics of the hairpin in presence of the applied 
force \cite{woodside,hanne,mossa}. These studies suggest that the average dissociation 
force increases logarithmically with the pulling speed \cite{mossa}.
  
\begin{figure}[t]
\includegraphics[width=2.6in]{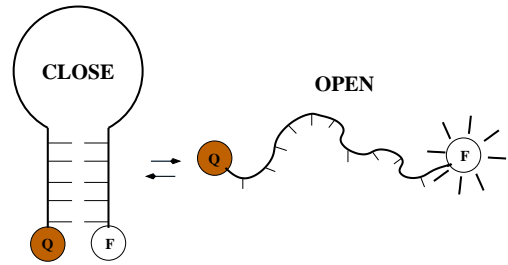}
\caption{The schematic representations of the ssDNA which can form a hairpin 
consisting of stem (complementary nucleotides at two ends) and a loop (made up of 
non-complementary bases). The hairpin fluctuates  between the close and open state. }
\label{fig-1}
\end{figure}

To have a better understanding of the biological processes, theoretical works 
\cite{montanari,zhou1,zhou2,Hugel} focused on simple models, which are either 
analytically solvable or accurate solution is possible through the extensive 
numerical simulations.  Theoretical analysis
of the elasticity of a polymer chain with hairpins as secondary structures \cite{montanari}
reproduces the experimental force-extension curve measured 
on the ssDNA chains, whose nucleotide bases are arranged in a  relatively random order.
The force induced transition in the hairpin is found to be of second order and 
characterized by a gradual decrease in the number of base pairs as the external 
force increases. Zhou et al. \cite{zhou1,zhou2} studied the secondary structure 
formation of the ssDNA (or RNA) both analytically as well as by the Monte Carlo simulations. 
They showed that the force induced transition is continuous from the hairpin-I 
(small base stacking interaction) to the coil, while a first order for the 
hairpin-II (large  base stacking interaction).  Hugel et al. \cite{Hugel} studied 
three different chains, namely ssDNA, poly vinylamine and peptide at very high 
force ($\sim$ 2 nN). 
At such a high force, conformational entropy does not play a significant role, 
therefore, zero temperature {\it ab initio} calculation has been applied to compare 
the experimental results. 

In many biological reactions, a slight change  in temperature causes a large change 
in the reaction coordinates \cite{danilow1}. Therefore, efforts of SMFS experiments
have now been shifted to study the effect of temperature keeping the applied force
constant. In this respect, Danilowicz {\it et al.} have measured  the critical force 
required for the unzipping of double stranded DNA (dsDNA) by varying temperature 
and determined the force-temperature phase diagram \cite{prentiss}. In another work 
\cite{danilow1}, it was shown that the 
elastic properties of ssDNA, which can form a hairpin, have significant temperature 
dependence. It was found that at the low force, the extension increases with temperature, 
whereas at the high force, it decreases with temperature. It was argued 
that the increase in the extension is the result of the disruption of hairpins. 
However, there is no clear understanding about the decrease in the extension with 
temperature, at high force. 
Moreover, the experimental force-temperature diagram of DNA hairpin remains elusive in 
the literature.

As pointed above, effect of loop length, sequence, nature of transition and the 
force-extension curve of DNA/RNA hairpin  are well studied at room temperature. 
In one of the earlier studies \cite{bonnet}, fluorescence and quencher were 
attached to the two ends of the stem with the assumption that the hairpin opens from 
the stem-end side \cite{Tinoco}. However, if the applied force is constant and 
temperature varies, the loop entropy may play a crucial role, whose 
effect to the best of our knowledge has not been addressed so far. In this paper, we focus 
mainly on two issues: (i) whether the force-temperature diagram of DNA hairpin differs significantly
with the dsDNA or not (Ref. \cite{km,bhat99,nsingh,maren2k2,rkp_prl} yielded qualitatively similar 
diagram), and (ii) precise effect of loop on the opening of DNA hairpin. 
In order to study such issues, we introduce a model of polymer with suitable constrain
to model the dsDNA and DNA hairpin and performed Langevin  Dynamics (LD) simulations 
\cite{Allen,Smith,janke} to obtain the thermodynamic observables, which have been 
discussed in section II. In section III, we validate the model, which describes some 
of the essential properties of DNA melting. In this section, we also show that the force 
induced melting of the hairpin differs significantly from the dsDNA. Section IV deals with
the semi-microscopic mechanism involved in the opening of the hairpin. Here, we discuss
various properties of loop {\it e.g.} entropy, length and stiffness and its consequence
on the opening.  We also revisited  lattice models in this section 
and discussed a possible reason for  discrepancies in the force-temperature
 diagram.
Finally in section V, we summarize our results and discuss the future perspective. 

\section{Model and Method}
The typical time scale involved in the hairpin fluctuations  varies
from $ns$ to $\mu s$, therefore, an all-atom simulation of the longer base sequences 
is not amenable computationally \cite{Hugel,pm}. In view of this, we adopt a minimal 
off-lattice coarse grained model of the DNA, where a bead represents a few bases
associated with sugar and  phosphate groups. 
In order to study the consequences of the loop entropy on the force-temperature diagram, 
we have considered a ssDNA which can form either a zipped conformation 
with no bubble/loop (Fig. 2a) or a hairpin (Fig. 2b ) in the chain depending on 
the base sequence.
For the zipped conformation, the first half of the chain (say made up of adenine {\bf C}) 
interacts with the complimentary other half of the chain (made up of thymine {\bf G}). 
The ground-state conformations resemble the zipped state of the dsDNA with no loop (Fig. 2a). 
However, if first few beads of the chain are complimentary to the last  beads and remaining 
nucleotides non-complimentary, we have the possibility 
of the formation of a hairpin (Fig. 2b) consisting of 
a stem and a loop at low temperature. In high force limit, the ssDNA acquires the 
conformation of the stretched state (Fig. 2c). Further refinements in the model requires inclusion 
of the excluded volume effect and proper base pairing. In addition,  
the elastic energy and the loop entropy, which play a crucial role in high force and  
high temperature regimes of the phase diagram, should also be incorporated in the 
description of the model.  The energy of the model system is defined as \cite{Li}

\begin{equation}
E = \sum_{i=1}^{N-1}k(d_{i,i+1}-d_0)^2+\sum_{i=1}^{N-2}\sum_{j>i+1}^{N}4(\frac{B}{d_{i,j}^{12}}  -\frac{A_{ij}}{d_{i,j}^6}) 
+ k_{\theta}(\theta - \theta_0)^2,
\end{equation}

\begin{figure}[t]
\includegraphics[width=2.6in]{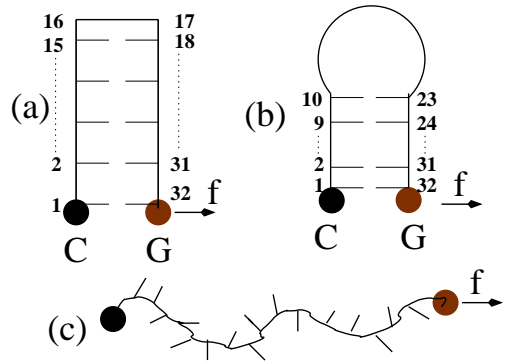}
\caption{The schematic representations of transformation  of the ssDNA to (a) the dsDNA, (b) a
hairpin consisting of a loop $\&$ stem and (c) the extended state.
The loop consists of non-complementary nucleotides of the stem.
}
\label{fig-2}
\end{figure}

\begin{table*}[t]
\centering
\tabcolsep 0.1pt
\caption {Native matrix elements ($A_{i,j}$) of Eq. (1) for  two conformational
possibilities: a dsDNA and a DNA hairpin}
\begin{tabular*}{\textwidth}{@{\extracolsep{\fill}}|c|c|c|c|c|c|c|c|c|c|c|c|c|c|c|c|}
\hline
 Case & A$_{1,32}$ & A$_{2,31}$ & A$_{3,30}$ & A$_{4,29}$ & A$_{5,28}$ & A$_{6,27}$ & A$_{7,26}$ & A$_{8,25}$ & A$_{9,24}$ & A$_{10,23}$ & A$_{11,22}$ & A$_{12,21}$ & A$_{13,20}$ & A$_{14,19}$ & A$_{15,18}$\\
\hline
dsDNA & 1 & 1 & 1 & 1 & 1 &  1 & 1 & 1 & 1 & 1 & 1 & 1 & 1 & 1 & 1\\
\hline
DNA hairpin & 1 & 1 & 1 & 1 & 1 & 1 & 1 & 1 & 1 & 0 & 0 & 0 & 0 & 0 & 0 \\
\hline
\end{tabular*}
\end{table*}

where $N$ is the number of beads. The distance between beads $d_{ij}$, is
defined as $|\vec r_i-\vec r_j|$, where $\vec r_i$ and $\vec r_j$ denote the position
of bead $i$ and $j$, respectively. The harmonic term with spring constant
$k$ couples the adjacent beads along the chain. We fixed $k = 100$, because the strength
of the covalent bond is almost 100 times stronger than the hydrogen bonding \cite{hbond}. 
The second term corresponds to the Lennard-Jones (LJ) potential. The  first term of LJ 
potential takes care of the "excluded volume effect", where we set $B = 1$. It should be 
noted that the hydrogen bonding is directional in nature \cite{bloom} and only one pairing 
is possible between two complementary bases. However, the model developed here is for 
the polymer, where a bead can interact with all the neighboring beads. In order to model 
DNA, one can assign the base pairing interaction $A_{ij} = 1$ for the native contacts (Table 1)
and 0 for the non-native ones  \cite{Li}, which corresponds to the Go model \cite{go,pablo}.
 By the 'native', we mean that the first base interacts with the $N^{th}$ (last one) base 
only and the second base interacts with the $(N-1)^{th}$ base and so on as shown in 
Fig. 2(a \& b). This ensures that the two complimentary bases can form at most one base 
pair. The remaining term of the Eq.(1) is the bending energy term, which is assigned to 
successive bonds in the loop only. Here, $k_{\theta}$ is the bending constant and $\theta$ 
is the angle between two consecutive bonds. $\theta_0$ is its equilibrium value. In our 
subsequent analysis, we will consider two cases namely $k_\theta$ = 0, which corresponds
to a loop made up of thymine  and  $k_\theta \ne 0$ that corresponds to adenine 
\cite{Goddard,ke,gk}. The Go model \cite{go,pablo} built on the assumption that the 
"energy" of each conformation is proportional to the number of native contacts, it poses 
and non-native contacts incur no energetic cost. By construction, the native state is the 
lowest energy conformation of the zipped state of DNA or DNA hairpin.  
Since, Go model exhibits a large energy gap between closed to open state and folds
rapidly to its ground state, therefore, it saves computational time.

It should be noted here that the model does not include the energetic of the slipped and
partially mismatched conformations. A more sophisticated model \cite{dg}, which includes
directionality of bases and takes care of proper base pairing (non-native interaction), 
gives rise to the existence
of intermediate states in the form of slipped and partially mismatched states. However,
for both the models, it was shown that the thermodynamic observables are almost
the same for the DNA unzipping \cite{dg}. In view of this, we prefer
native base pairing interaction in the present model. The parameter $d_0 (=1.12)$ 
corresponds
to the equilibrium distance in the harmonic potential, which is close to the equilibrium
position of the average LJ potential. In the Hamiltonian (Eq. (1)), we use dimensionless
distances and energy parameters. The major advantage of this model is that the ground
state conformation is known. Therefore, equilibration is not an issue here, if
one wants to study the dynamics steered by force at low $T$ \cite{Li}.
We obtained the dynamics by using the following Langevin equation
\cite{Allen,Smith,Li,MSLi_BJ07}
\begin{equation}
m\frac{d^2r}{dt^2} = -{\zeta}\frac{dr}{dt}+F_c+\Gamma
\end{equation}
where $m$ and $\zeta$ are the mass of a bead and the friction
coefficient, respectively. Here, $F_c$ is defined as $-\frac{dE}{dr}$ and
the random force $\Gamma$ is a white noise \cite{Smith}
{\it i.e.} $<{\Gamma(t)\Gamma(t')}>=2\zeta T\delta(t-t')$.
The choice of this dynamics keeps $T$ constant throughout the simulation for a given
$f$. The equation of motion is integrated using the $6^{th}$ order predictor
corrector algorithm with a time step $\delta t$ = 0.025 \cite{Smith}.
 We add an energy $-\vec{f}.\vec{x}$ to the total energy
of the system given by Eq. (1).

We calculate the thermodynamic quantities after the equilibration using the native state 
as a starting configuration. The equilibration has been checked  by calculating the 
auto-correlation function of any observable $q$, which is defined as \cite{box,berg} 
\begin{equation} 
\hat{S}_q(t) = \frac{\langle q(0) q(t) \rangle - \langle q (0) \rangle^2}{\langle q^2(0) \rangle - 
\langle q(0) \rangle^2},
\end{equation} 
Here, $q$ can be the end-to-end distance square or the radius of gyration square.
The asymptotic behavior of $\hat{S}_q(t)$ for large $t$ is 

\begin{equation} 
\hat{S}_q(t) \sim \exp(-\frac{t}{\tau_{exp}}) 
\end{equation} 

where $\tau_{exp}$ is so called the (exponential) auto-correlation time. 
In general, the equilibration can be achieved after $2 \tau_{exp}$ \cite{box,berg}. 
In our simulation, we have chosen the equilibration time which is ten times 
more than the $\tau_{exp}$. The data has been sampled over four times 
of the equilibration time.  We have used $2\times10^9$ time
steps out of which the first $ 5\times 10^8$ steps have not been taken in the
averaging.  The results are averaged over many trajectories which are almost 
the same within the standard deviation.
We also notice that at low $T$, it is difficult to achieve the
equilibrium and the applied force probes the local minima instead of
global minima.

\section {Results}
\subsection{Equilibrium property of dsdNA and DNA hairpin at $f=0$}

Before we discuss the underlying physics behind the force-induced transition,
we would like first to validate our model based on Eq. (1) and show that the
model does include some of the essential features of DNA and melting show the 
two state behavior. The reaction co-ordinate {\it i.e.} absolute value of 
end-to-end distance ($|x|$) of DNA hairpin and dsDNA have been plotted as a  
function of temperature in Fig. 3. At low temperature,  the thermal energy is 
too small to overcome the binding energy and DNA hairpin(dsDNA) remains
 in the closed state(zipped state), and end-to-end distance remains quite 
small. At high temperature,
the thermal energy is quite large compare to its binding energy  and the chain 
acquires the conformations corresponding to the swollen state (open state), where 
$|x|$ scales with $N^{\nu}$. Here $\nu$ is the end-to-end distance exponent and 
its value is given by $\frac{3}{d+2}$ \cite{degennes}. The variation $\frac{|x|}{N^\nu}$ 
can be fitted by the sigmoidal distribution \cite{kenward} with the melting temperature 
$T_m$ = 0.21 \& 0.23 for DNA hairpin and dsDNA, respectively. The melting temperature 
can also be obtained by monitoring the energy fluctuation ($\Delta E$) or the 
specific heat ($C$) with temperature, which are given by the following relations 
\cite{Li}:

\begin{eqnarray}
<\Delta E> & = & <E^2>-<E>^2  \\
C & = & \frac{<\Delta E>}{T^2}.
\end{eqnarray}

The peak in the specific heat curve gives the melting temperature that matches with 
the one obtained by the sigmoidal distribution of end-to-end distance. This shows 
that for a small system transition is well described by the two state model.

\begin{figure}{t}
\includegraphics[width=3.5in]{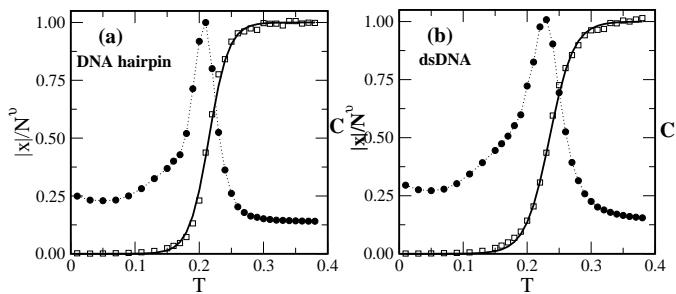}
\caption{ Variation of normalized extension(open square) and specific heat(C)
 (filled circle) with temperature. Solid line corresponds to the sigmoidal fit.
(a) for DNA hairpin case (b) for dsDNA case.}

\end{figure}

\subsection{Force induced melting}
In  many biological reactions involve a large conformational change
in the  mechanical reaction coordinate {\it i.e.} $|x|$ or the number of 
native contacts, that can be used to follow the progress of the reaction 
\cite{busta}. As discussed above, the two state model describes these processes 
quite effectively in the absence of force. The applied force ''tilts” the 
free energy surface along the reaction coordinate by an amount linearly 
dependent on the reaction coordinates \cite{busta,kumarphys}. One of the notable aspects 
of the force experiments on single biomolecules is that $|x|$ is directly 
measurable or controlled by the instrumentation, therefore, $|x|$ becomes a 
natural reaction coordinate for describing the mechanical unzipping.

The critical unzipping force, which is a measure of the stability of the DNA hairpin (dsDNA) 
compared to the open state, has been determined by using Eqs. 5 and 6 as a function of 
temperature. The phase boundary in the force-temperature diagrams (Fig. 4a \& b) 
separate the regions where DNA hairpin (dsDNA) exists in a closed state (zipped 
for dsDNA) from the region where it exists in the open state (unzipped state).
It is evident from the plots (Fig. 4 a \& b) that the melting temperature decreases 
with the applied force in accordance with the earlier studies \cite{bhat99,nsingh,maren2k2,rkp_prl}. 
The peak position of $\Delta E$ and $C$ coincides, and the peak height increases 
with the chain length ($26$, $32$ and $42$, where the ratio of stem to loop has been kept 
constant). However, the transition temperature (melting temperature) remains almost the 
same in the entire range of $f$. Since, experiments on hairpin are mostly confined to 
small chains \cite{wallace,bonnet,Goddard,woodside}, here, we shall confine our-self to 
the chain of $32$ beads to study the phase diagram and related issues.

\begin{figure*}[t]
\begin{center}
\includegraphics[width=4.5in]{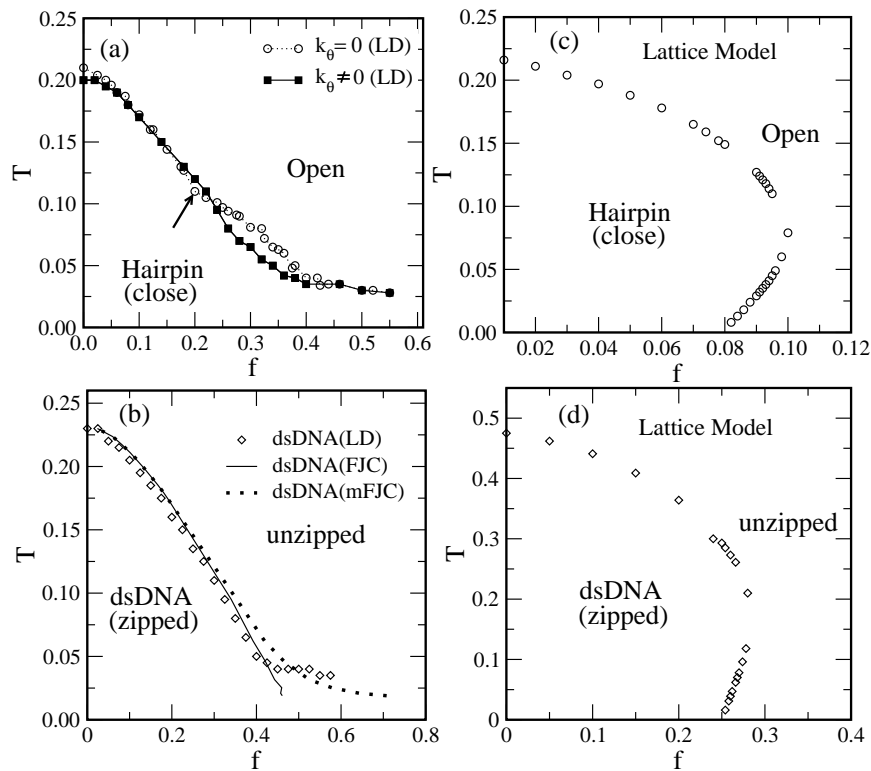}
\end{center}
\caption{(a) $f-T$ diagrams for a DNA hairpin obtained from LD simulations with and without 
the bending energy term. The arrow indicates the change in the slope which vanishes for 
$k_{\theta} = 20$;
(b) $f-T$ diagrams for the dsDNA using LD simulations. The phase diagram is also compared with 
the FJC and the mFJC models in the reduced unit. Our results are in good agreement with the 
FJC and the
mFJC. The deviation at high force is because of the elastic energy, which is included in the mFJC; 
(c \& d)  The phase diagrams for the DNA hairpin and the dsDNA using lattice model, which are 
qualitatively similar to each other. Clear differences between simulation and lattice model 
are  visible at intermediate and high force regime of the phase diagram.}
\label{fig-3}
\end{figure*}

For the sake of comparison, in Fig.4c and 4d, we also show the phase diagrams of DNA 
hairpin (stem length =3, loop length =12 and bond length =2) and dsDNA of 9 base pairs (bps)
on the cubic lattice using exact enumeration technique\cite{km}. 
The force-temperature diagrams were found to be in qualitative agreement with earlier 
theoretical predictions \cite{bhat99,maren2k2,rkp_prl,gk}. Despite the simplicity involved 
in the lattice model, it provides a deeper insight in the mechanism involved in the force 
induced transitions \cite{kumarphys}. A self attracting self self-avoiding walk (SASAW) 
on appropriate lattice (here cubic lattice) may be used to  model a DNA hairpin 
(stem and loop) and dsDNA (zipped). The base pairing is assigned ($\epsilon = -1$), when 
the two complimentary nucleotides are on the nearest neighbor. The nearest neighbor 
interaction mimics the short range nature of the hydrogen bonds, which are qualitatively 
similar to the model adopted here.

It is pertinent to mention here that for the small base pair sequences ($ < 100$), 
the differential melting curve of the dsDNA shows a single peak \cite{Wartel}, indicating 
a sudden unbinding of two strands. Since for the short dsDNA, the entropy contributions of 
spontaneous bubbles are not very important as  loops are rare, therefore, transition is
well described by the two state model \cite{Wartel}. One would expect that the force-temperature 
diagram of the short dsDNA presented here should then match with the two state model.  
The free energy of the unzipped chain ($g_u$) under the applied force can be obtained through 
the freely jointed chain (FJC) model \cite{strick}
\begin{equation}
g_u^{FJC}=\frac{l}{b}k_bT\ln[\frac{k_bT}{f b}\sinh(\frac{f b}{k_bT})],
\end{equation}
which can be compared with the bound state energy ($g_b$) of the dsDNA to get the
force-temperature diagram, which is shown in Fig.4b. Here, $b$ and $l$ are the Kuhn 
length and  bond length of the chain, respectively. In the reduced unit for the dsDNA,
one can notice a nice agreement between the off-lattice simulation presented here and the 
two state model (Eq. 7) over a wide range of $f$ and $T$. 
However, at low $T$ and high $f$, the phase boundary deviates from the 
FJC model. One should recall that the first term of the Hamiltonian defined in 
Eq. (1) may induce the possibility of the stretched state at high force, where the 
bond length may exceed than its equilibrium  value. In fact in the FJC 
model,  the bond length is taken as a constant. It should be emphasized here that the 
modified freely jointed chain (mFJC) model does include the possibility of the stretching 
of  bonds in its description \cite{prentiss,Dessinges}. Therefore, if this deviation is because of 
the elastic energy then the two state model based on the mFJC model should show a similar 
behavior as seen in the simulation. The free energy of the unbound state of the mFJC 
may be obtained by including $\frac{l}{2b}\frac{(f{\ell})^2}{k_bT}$ term in Eq. (5) 
\cite{prentiss,Dessinges}, where $\ell$ is the increase in the bond length. The corresponding 
force-temperature phase diagram in the reduced unit shown in Fig.4b is in excellent 
agreement with the simulation in the entire range of $f$ and $T$. 

\begin{figure}[b]
\begin{center}
\includegraphics[width=2.in]{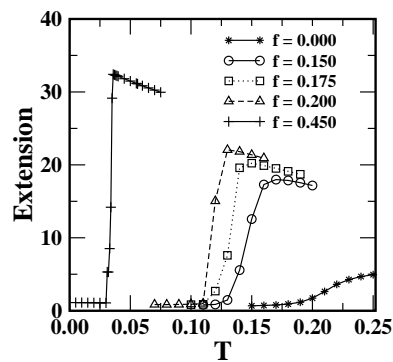}
\end{center}
\caption{Temperature-extension curves at different values of the force.}
\end{figure}

The off-lattice simulation performed here, clearly shows the 
effect of loop of the hairpin on the melting profile. The phase diagram of the hairpin
(Fig.4a) differs significantly from the dsDNA (Fig.4b) as well as deviates considerably 
with the counter lattice model of the hairpin (Fig.4c). Two major differences can be noticed 
from the figures: (i) a change in the slope at the intermediate value of force for the 
hairpin (Fig. 4a), which is absent (Fig.4b) in the case of dsDNA (no loop) as well as in 
the  previous theoretical models \cite{bhat99,nsingh,maren2k2,rkp_prl}; 
(ii) an abrupt increase in the force for both the hairpin and the dsDNA at low temperature (Fig.4a 
and 4b), whereas the lattice model and other theoretical studies \cite{bhat99,nsingh,maren2k2,rkp_prl} 
show the re-entrance for both hairpin and DNA (Fig. 4c and 4d).

To have further understanding of it, we have monitored the reaction coordinate 
($|x|$) of the hairpin with temperature at different 
values of $f$ (Fig.5). At low $f$, transition is weak due to the finite width of the
melting profile, which decreases with $f$. At high $f$, however, it shows a strong 
first order characteristic. At high force, the non-monotonic behavior of the extension 
with temperature as observed in a recent experiment \cite{danilow1} is also evident 
from Fig.5. As $T$ increases, the chain acquires the stretched state (Fig.2c) because of
the applied force. A further rise in $T$ results an increase in the contribution of 
the configurational entropy of the chain. The applied force is not enough to hold the stretched 
conformation and thus the extension falls \cite{km,kumar_prl}. At intermediate $f$ 
$(\sim 0.19)$ where the change in the slope is observed (Fig. 4a),  we find that the entropic ($T\;dS$)
and the force ($f\;dx$) contributions to the free energy are nearly the same.

At constant temperature, by varying $f$, the system attains the extended states from 
the closed state of the hairpin. With further rise of $f$, one finds the stretched state.
{\it i.e} the extension approaches the contour length of the $ssDNA$.
At high $T$,  when $f$ increases hairpin goes to the extended state.
At the same T, if we decrease the applied force, the system retraces the path to the
closed state without any hysteresis. Because of high entropy, it is
possible that monomers of two segments of strand come close to each other 
and re-zipping takes place (Fig. 7a). As we decrease the $T$ below $0.15$, by increasing $f$, 
hairpin acquires the stretched state. Interestingly, now it doesn't retrace the path
if $f$ decreases at that $T$ (Fig. 7b). This is the clear signature of hysteresis. 
It is because of the contribution of entropy, which is not sufficient enough to bring  
two ends close to each other so that re-zipping can take place. The hysteresis has been 
measured recently in unzipping and re-zipping of DNA \cite{Hatch}. It was found that
the area of hysteresis loop increases with decreasing temperature, which is also consistent
with our findings (Fig. 6b).

\begin{figure}[t]
\begin{center}
\includegraphics[width=3.5in]{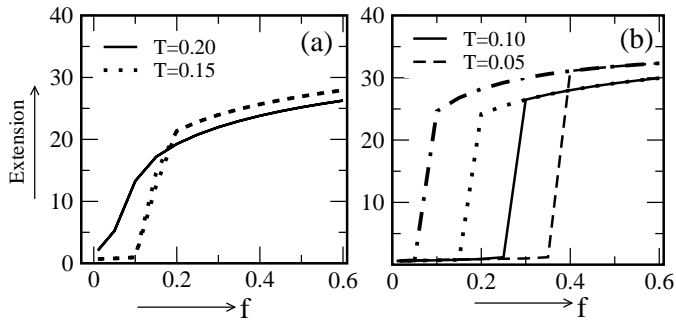}
\end{center}
\caption{Variation of extension (up to contour length) with $f$ at $T > 0.15$. The path retraces and
there is no signature of hysteresis; (b) same as (a), but at $T < 0.15$. Path
does not retrace with decreasing $f$ and a force is required to re-zip, which
is a signature of hysteresis (dotted and dotted-dashed line)
.}
\end{figure}


\section{Effect of loop on the opening of hairpin}

The thermodynamics of DNA melting is well studied using the following equation \cite{woodson}:
\begin{equation}
     \Delta G = \Delta H - T \Delta S_{stem}-T \Delta S_{loop},   
\end{equation}                                     
where $G, H,  S_{stem}$ and $S_{loop}$  are the free energy, enthalpy, 
entropy associated with stem and loop, respectively. To determine the entropic
force of the loop,  we fix the stem length and vary the loop length from 0 to
N/2 at fixed T. Since the contribution of first two term of Eq. 8 remain constant as 
stem length is kept fixed, the decrease in the applied force is the result of the
loop entropy. In Fig. 7, we plot the force as a function of loop length at two different
temperatures. At high temperature, the loop entropy reduces the applied force (Fig. 7a), and
hence, there is a decreases in the force, which is in accordance with recent experiment 
\cite{woodside}. However, at low temperature, the applied force remain constant (Fig. 7b) indicating 
that the loop entropy does not play any role in the opening of the hairpin and opening 
is mainly force driven. This is because at low temperature, applied  force 
probes the local minima and hence remain independent of global minima. However,
rezipping force does depend on the loop length.

\begin{figure}[h]
\includegraphics[width=3.0in]{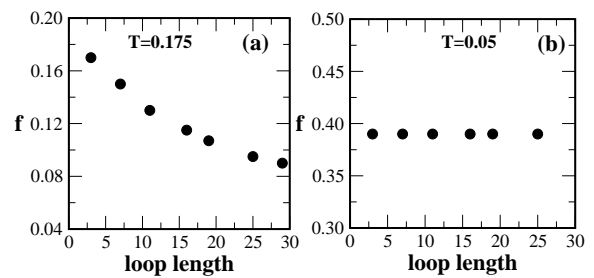}
\caption{Variation of the applied force with loop length at high and low temperatures.}
\end{figure}

\begin{figure*}
\includegraphics[width=5.in]{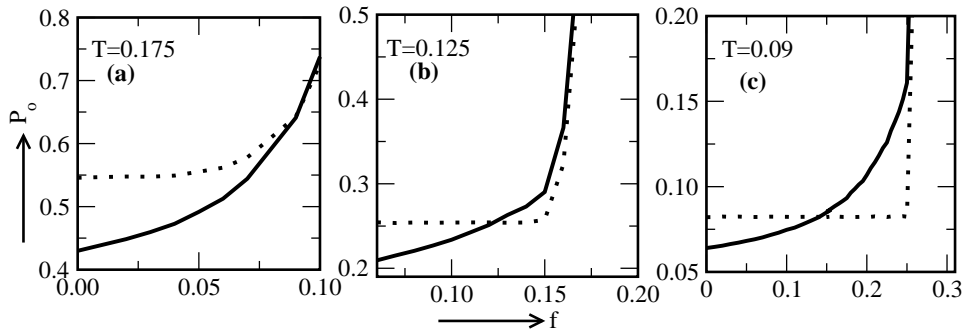}
\caption{(a) Comparison of the probability of opening ($P_o$) of the stem-end
(solid line) and the loop-end (dotted line) with the force for different temperatures.
Near the transition point (Fig. 4a), the probability of
opening is higher for the loop-end; (b) The two probabilities are comparable, and hence, the
hairpin can open from both sides; (c) It reflects that opening is always from the stem-end side.}
\label{fig-5}
\end{figure*}

In order to substantiate our findings, we tracked  the first
($1-32$, stem-end) and the last ($10-23$, loop-end) base pairs of the stem (Fig. 2b)
in different regimes of the $f-T$ diagram. In Fig. 8, we show the probability
of opening ($P_o$) of the stem-end and the loop-end base pairs with $f$ at a given $T$.
Near the phase boundary, at high $T$, transition from the hairpin to the coil state is entropy
driven as discussed above. This is apparent from Fig. 8a, where $P_o$ for the loop-end 
is higher than that of the stem-end. This indicates that in the opening of the hairpin, 
the loop contributes significantly compared to the applied force. At intermediate $T\; 
(\sim 0.12)$, $P_o$  of the stem-end  and the loop-end are comparable, reflecting 
that the hairpin can open from any sides (Fig. 8b). However, at low $T$, the dominant 
contribution in the opening of hairpin comes from the stem-end side and the 
transition is force driven (Fig. 8c).

It remains a matter of quest that why the exact solution of 
the lattice models of hairpin \cite{km}  could not exhibit this feature.  This 
prompted us to revisit the lattice model studied in Ref. \cite{km}. We calculated 
$P_o$ of the stem-end and the loop-end for the lattice model of the hairpin. At high $T$, 
opening of the hairpin, the dominant contribution comes from the stem side of the 
hairpin. This should not be taken as surprise as lattice models do not take the loop entropy 
properly in their description. In fact, the discrete nature of the lattice 
(co-ordination number) reduces the loop entropy. The model studied here is 
the off-lattice one, where the loop entropy has been taken properly into account assuming 
that ssDNA is well described by the FJC model \cite{smith,Dessinges}.
 Therefore, 
this behavior occurs in our model system.

If the change in the slope (Fig. 4a) is because of the loop entropy (underestimated 
in the lattice models), then the change in the slope should also vanish even in
 the 
off-lattice model, if the loop entropy is reduced somehow. This can be achieved 
by putting $k_{\theta} \ne 0$ in the loop [Eq. (1)]. We repeated the simulation with the 
bending constant $k_{\theta}$ in between 10 to 20, which have been used in earlier 
simulations \cite{kouza,Noy,alex}. Because of this term, the change in the slope 
seen at intermediate value of $f$ vanishes (Fig. 4a). The most 
striking observation here is that the probability distribution analysis shows 
that the opening of hairpin is contributed by the stem-end side, even at high 
temperature. 

\section{Conclusions}
In conclusion, we have studied a simple model of a polymer, which can form a 
dsDNA or a DNA hairpin depending on the interaction matrix given in Table 1. 
The model includes the possibility of the stretched state in its description. 
Since, simulation is performed on the off-lattice, therefore, the loop entropy 
has been taken properly into account in case of the hairpin. The $f-T$ diagrams 
of the hairpin differs significantly from the dsDNA as well as previous studies 
\cite{km,bhat99,nsingh,maren2k2,rkp_prl}.

We show that the loop entropy plays an important role, which has been 
underestimated in previous studies. This conclusion is based on the 
probability analysis of the extreme bases of the stem (loop-end \& stem-end), 
which rendered the semi-microscopic mechanism involved in the force induced 
melting. Our results provide support for the existence of the apparent 
change in the slope in the phase diagram. In the intermediate range of $f$ and 
$T$, a chain can open from both sides, reflecting that the entropic and force 
contributions are nearly the same, which may be verified experimentally. 
At high $T$, the transition is entropy dominated and the loop contributes in the
opening of hairpin (Fig. 6a). At low $T$, it is force driven and it opens from the 
stem-end side (Fig. 6b). It should be noted here that, in the description of the
lattice model, the critical force for DNA hairpin near $T = 0$  differs 
significantly with dsDNA (Fig. 4c \& d), where as in the simulation the unzipping
forces for both the cases were found nearly the same (Fig. 4a $\&$ b). 
This should not be taken 
as a surprise because  in the exact enumeration, one has the complete information 
of the partition function and
hence the global minimum can be obtained to calculate the critical force 
at low T.  However, in present simulation, we follow the experiments and 
report the dynamical force required to unzip the chain in Figs. 4 a $\&$ b and Fig. 7 
\cite{Hatch,PNAS}, which probes the local minima at low T. 
This force may differs with the critical force at low $T$. For example, for the 50000
bases $\lambda-$DNA, the critical force was found to be 15.5 pN to overcome the 
energy barrier $ \sim 3000 K_BT $. Full opening of the molecule never happens
in the experiment as time required to open the barrier is quite large. Thus a
larger force (17 pN) than its critical force is require to open a finite 
fraction \cite{EPJE_cocco} of the chain.    

It is important to mention here that a  change in the phase boundary has also 
been observed in the force induced unzipping of $\lambda$-phage dsDNA 
consisting of approximately 1500 base pairs. For such a long chain, 
differential melting curve shows several peaks corresponding to the partial 
melting of dsDNA, which form spontaneous bubbles in the chain \cite{Wartel}. This 
suggests that for a longer chain, formation of spontaneous bubble should be 
incorporated in the model. Using Peyrard-Bishop (PBD) model \cite{Peyrard}, 
Voulgarakis {\it et al.} \cite{Voulgarakis} probed the 
mechanical unzipping of dsDNA with bubble. One of the important results
they derived, is that the single strand as it extends between the dsDNA
causes a decrease in the measured force but the $f-T$ diagram remains  
significantly different than the experiment.  Recalling that in 
the PBD model, a bead can move only in one-dimension, and hence the entropy 
of the loop is also underestimated here, which may be the reason for such 
difference. Therefore, at this stage of time a simulation of longer base 
sequences is required, which should have spontaneous bubbles in its 
description with proper entropy to settle this issue.

The another interesting finding of our studies is the absence of re-entrance 
at low temperature. The basic assumption behind the theoretical models or 
phenomenological argument is that the bond length in the model system is  
a constant. However, recent experiments \cite{prentiss,ke} suggests that there is a net 
increase in the bond length at high force. Consequently model studied here
incorporates the simplest form of the elastic energy ({\it i.e.} Gaussian 
spring). We have kept the spring constant quite large in our 
simulation to model the covalent bond. The experimental force-temperature 
diagram of $\lambda-$phage DNA \cite{prentiss} shows the 
similar behavior at high force.
The observed decrease in the slope at low temperature in the experiment is attributed to a
thermally induced change in the dsDNA conformation \cite{prentiss}. However, present simulation
(DNA hairpin and dsDNA) along with the two state model based on the mFJC suggests 
that such deviation may be because of 
the elastic energy. In earlier studies, either this energy has not been included in the models 
\cite{Peyrard,Voulgarakis,kafri} or the bond length has been taken as  a constant 
\cite{bhat99,nsingh,maren2k2,km}, thus precluding the existence of the stretched state with 
the increased bond length in both cases (dsDNA and hairpin). Consequently, these
models could not describe this feature of abrupt increase in the force at low $T$.
In future  theoretical studies, one may use finite extensible nonlinear elastic (FENE) 
potential \cite{FENE1} and proper loop entropy {\it e.g} one used in Ref. \cite{bundschuh} to have further
understanding of the force-temperature diagram of a longer dsDNA. 

\section{acknowledgments}
We would like to thank S.
M. Bhattacharjee, M. Prentiss, D. Mukamel, D. Marenduzzo and  Y. Kafri for their comments 
and suggestions. Financial supports from the CSIR, DST, India, Ministry of Science and 
Informatics in Poland (grant No 202-204-234), and the generous computer support from MPIPKS, 
Dresden, Germany are acknowledged.

\end{document}